# Control of electronic band profiles through depletion layer engineering in core-shell nanocrystals

Michele Ghini[†,1,2] Nicola Curreli[†,3*] Matteo B. Lodi,[4] Nicolò Petrini,[3] Mengjiao Wang,[3,5] Mirko Prato,[5] Alessandro Fanti,[4] Liberato Manna,[1] Ilka Kriegel[3*]

1 - Department of Nanochemistry, Istituto Italiano di Tecnologia, via Morego 30, 16163 Genova, Italy
2 - Dipartimento di Chimica e Chimica Industriale, Università degli Studi di Genova, Via Dodecaneso 31, 16146 Genova, Italy
3 - Functional Nanosystems, Istituto Italiano di Tecnologia (IIT), via Morego 30, 16163 Genova, Italy
4 - Department of Electrical and Electronic Engineering, University of Cagliari, via Marengo 2, 09123, Cagliari, Italy
5 - Materials Characterization Facility, Istituto Italiano di Tecnologia, Via Morego 30, 16163, Genova, Italy
† - These authors contributed equally to this work.

## Abstract:

The precise understanding of depletion layers is of major importance to control the optical and electronic properties of metal oxide (MO) nanocrystals (NCs). Depletion layers typically form through the electronic depletion of the near surface region due to the presence of surface states. In this article, we show that depletion layer engineering is possible beyond surface states and that it is the main mechanism of photodoping of MO NCs. Exemplified by the case study of $Sn:In_2O_3/In_2O_3$ core-shell NCs with varying shell thickness, we show that the introduction of more than one electronic interface induces a double bending of the electronic bands accompanied by a distinct carrier density profile. This induces the formation of a depleted region between the shell and the surface of the nanoparticle effectively separating the particle into three distinct electronic regions: a core region with specific carrier density, a transition region with order of magnitude lower carrier density and an undoped depletion region. Notably, the electronic band profile does not coincide with the physical core-shell boundaries. We found that the light-induced depletion layer modulation and bending of the bands close to the surface of the nanocrystal is the main mechanism responsible for the storage of extra electrons after photodoping in MO NCs. We support our results by a combined experimental and theoretical approach in which we compare numerical simulations with empirical modelling and experiments. This allows us not only to extract the main mechanism of photodoping in MO NCs, a process so far not understood electronically, but also to engineer the charge storage capability of MO NCs after photodoping. Our results are transferable to other core-shell and core-multishell systems as well, opening up a novel direction to control the optoelectronic properties of nanoscale MOs by designing their energetic band profiles through depletion layer engineering.

## Introduction

Doped metal oxide (MO) nanocrystals (NCs) are gaining the attention of the scientific community thanks to their unique properties, such as high electron mobility,[1] the tunability of their carrier density over several orders of magnitude,[2] chemical stability,[3] and low toxicity,[3] as well as suitable operating temperature,[1] which makes them appropriate for a large plethora of applications, ranging from nanoelectronics and plasmonics to the next-generation energy storage.[3–11] In doped MO NCs, surface states, such as surface trap states, defects, vacancies, as well as surface ligands and other bound molecules induce Fermi level pinning causing an upward bending of the energetic bands.[2,4,12–18] The spatially varying conduction band translates into a gradient in the carrier density ($n_e$), sufficient to suppress entirely the metallic behaviour of carriers close to the nanocrystal surface. This depletion region effectively acts as a dielectric.[2,12,17,19] Hence, the homogeneous flat-band model, which neglects Fermi level pinning, is not sufficient to accurately describe the behaviour of free carriers in MO NCs, as introduced by other groups.[2,17,19,20] In fact, the depletion layer formation considerably affects the conductivity of NC films and their plasmonic behaviour, with direct implications on the electric field enhancement, the localized surface plasmon resonance (LSPR) modulation and its sensitivity to the surroundings.[2,17,19] Furthermore, the presence of a surface depletion region induces an important alteration to the particle dielectric function.[2]

Given the strong impact of depletion layers on the optoelectronic properties of nanoscale oxide materials, in this work, we aimed at exploiting the depletion layer formation to control energetic band profiles as a means to understand and improve material characteristics. We explore depletion layer engineering beyond surface states by introducing additional electronic interfaces and by dynamically modulating the carrier density via post-synthetic approaches. We experimentally exemplify this scheme with Sn-doped Indium Oxide (ITO)-$In_2O_3$ core-shell NCs and the fine-tuning of the shell thickness ($t_s$) as well as capacitive charge injection with light (*i.e.*, photodoping). Numerical simulations on both cases serve as a framework to describe in detail the nanoscale evolution of their electronic structure supported by an empirical model that describes the experimental optical properties of all NCs before and after photodoping. The empirical fit model together with electron counting experiments support the band structure calculations well. Through this combined theoretical and experimental work, we unveil that double band bending is a key characteristic of ITO-$In_2O_3$ core-shell NCs, describing well also the dynamic introduction of extra electrons via photodoping, a process not fully explained yet. We found that the photo-induced band bending results in an increase in $n_e$ predominantly in shell, contradicting the previously reported explanation of a uniform rise of the flatband Fermi level as main mechanism for photodoping.[11,16,21,22] Furthermore, the observed band bending supports charge separation towards the NC interface and avoids possible recombination. We finally exploit depletion layer engineering to improve the capacitive charging process in doped metal oxide nanocrystals upon photodoping, resulting in an accumulation of more than 600 stored charges per nanocrystal of the same size.

## Discussion

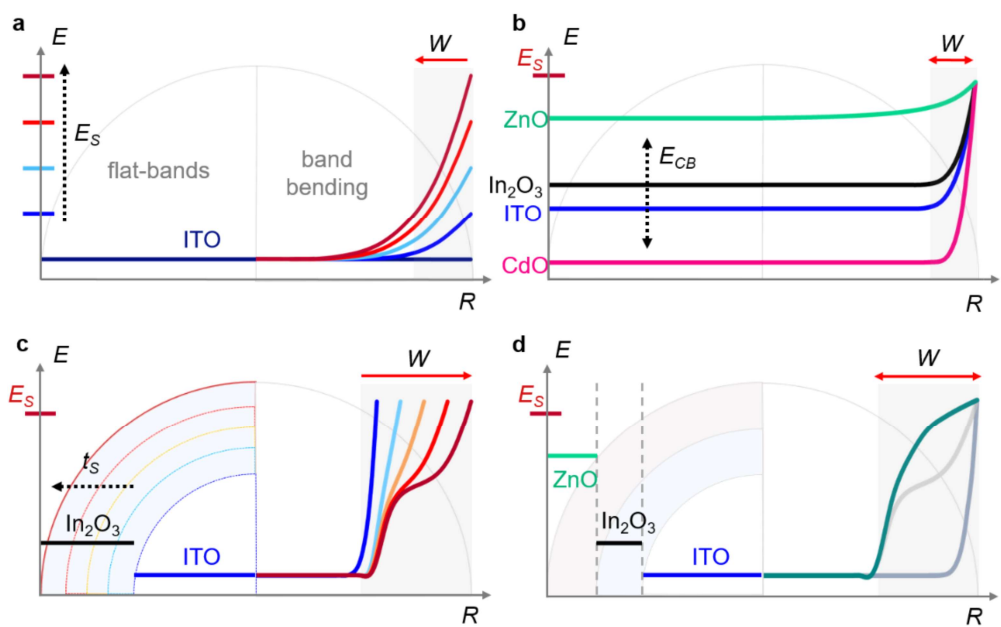

**Fig. 1 | Depletion layer engineering of metal oxide NCs via tuning of structural and electrical properties.** The morphology of the NC is illustrated by a semi-circle with its extension $R$ in the abscissa (in nm) and the energy in the ordinate (in eV). In each panel, the left side of the sphere represents the flat-band potentials in the non-equilibrium condition, while the right side reports the band bending due to the Fermi level pinning. Band profiles were calculated by numerically solving the Poisson's equation. **a,** Tuning of surface state potentials ($E_S$). Increasing the $E_S$ results in the expansion of the depletion width $W$. **b,** Impact of different materials on $W$ at fixed $E_S$. c, Expansion of $W$ and double bending of the depletion layer in a core-shell structure of ITO-$In_2O_3$ with a core radius ($R_{core}$) of 5.5 nm and varying shell thickness ($t_s = 0, 1, 2, 3, 4$ nm, i.e., blue, light blue, orange, red and dark red, respectively). **d,** Multiple shell system by combining an ITO core ($R_{core}$= 5.5 nm) with a $In_2O_3$ and ZnO shell with total radius $R$ = 9.5 nm. The band shows a complex profile with a triple bending (green curve). The grey curves illustrate the previously reported case of a uniform ITO NC (dark grey) and an ITO-$In_2O_3$ core-shell NC (light grey) with total radius $R$ = 9.5 nm for comparison.

We performed numerical simulations based on the solution of Poisson's equation[2] within the parabolic band approximation to illustrate the band structure of NCs and their depletion layer formation (extended details on the calculations are reported in the Supplementary Information).[23,24] Here, the depletion layer is defined as the region of the NC where $n_e$ drops below $10^{26}$ m$^{-3}$ (threshold value at which we can detect plasmonic features).[19] In **Fig. 1** we show the spatially dependent profile of the conduction band as a result of the upward band bending and its effect on the depletion layer width ($W$) for different parameters, such as surface potentials ($E_S$) (**Fig. 1a**), different materials (**Fig. 1b**) and the introduction of additional electronic interfaces (**Fig. 1c** and **Fig. 1d**). In the first case, we consider the effect of surface states on the depletion layer formation. The effect of Fermi level pinning is modeled by a fixed surface potential ($E_S = 0, 0.5, 1, 1.5, 2\ eV$), from which the band bending profile is derived. An ITO/surface electronic interface is formed. The value of $E_S$ can be found experimentally and it is a peculiar parameter for each material interface. It depends on several factors, such as specific densities of trap states, presence of defects and vacancies as well as surface ligands.[2] For increasing $E_S$, we observe an increase in the depletion layer width, which affects a larger fraction of the NC volume (**Fig. 1a**). These results indicate the importance of surface control to engineer the band structure of NCs. **Fig. 1b** reports the effects of changing the composition of the NC while keeping $E_S$ at a fixed energy. The choice of material, the elemental composition, permittivity ($\varepsilon$), bandgap energy ($E_g$) and the control over doping levels are of fundamental importance. Different materials, as in this case ZnO, In$_2$O$_3$, ITO and CdO, have a specific impact on the band bending, showing that $W$ is a unique feature of each system. Another powerful parameter to control the depletion layer and the energy level profile is the introduction of additional electronic interfaces beyond the surface of the nanoparticle. One example is the ITO-In$_2$O$_3$ core-shell nanocrystal system (**Fig. 1c**).

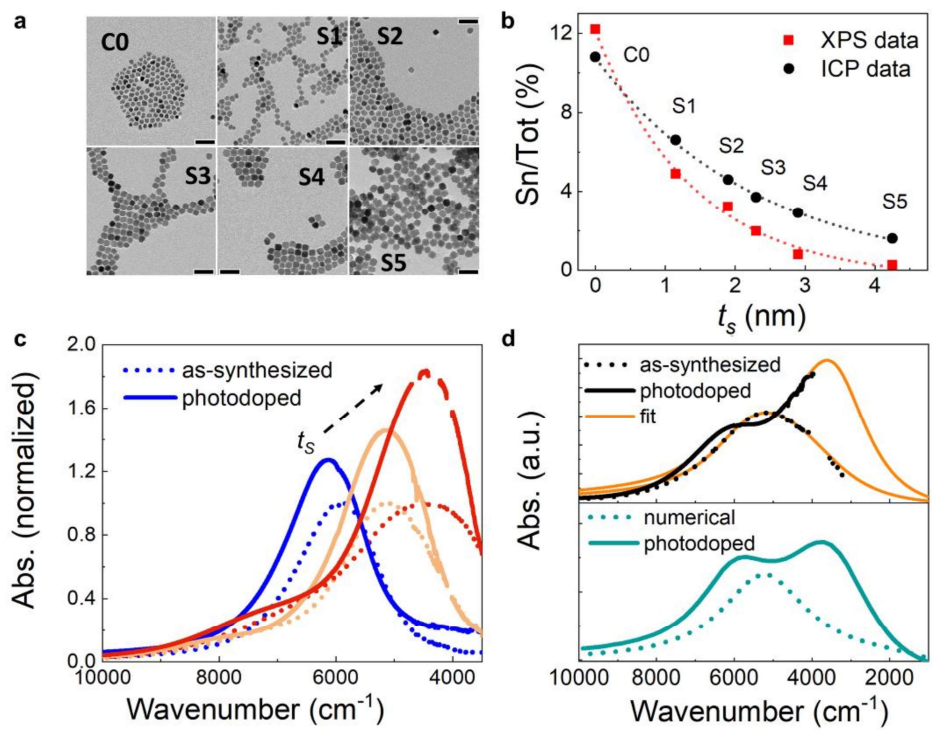

**Fig. 2 | ITO-In$_2$O$_3$ core-shell colloidal nanocrystals. a,** TEM images of the nanocrystals at different stages of the synthesis. Starting from an aliquot of the core (C0), layers of In$_2$O$_3$ progressively form a thicker shell around the ITO nuclei (S1-S5). Scale bar is 50 nm for each panel. **b,** Comparison between the Sn-dopant concentrations obtained from inductively coupled plasma mass spectrometry (ICP-OES) and X-ray Photoelectron Spectroscopy (XPS) as a function of shell thickness ($t_s$). The higher Sn/Tot values of the volume sensitive ICP-OES measurements with respect to the surface sensitive XPS measurements indicate that Sn atoms remain segregated in the core of the NCs. Dashed lines are to guide the eye. **c,** Normalized absorbance of typical ITO-In$_2$O$_3$ core-shell samples with increasing shell size, in as-synthesized (dotted lines) and photodoped (continues lines) cases, with a $t_s$ of 0.15 nm (blue), 1.85 nm (orange), and 2.3 nm (red). Growing an undoped shell continuously redshifts the energy of the localized surface plasmon resonance (LSPR) peak in the NIR. After 20 min of UV exposure, the intensity of the LSPR peak increases significantly, reaching values up to almost double its initial absorbance. **d,** Top panel: Absorbance of sample S5 ($t_s = 4.25$ nm), as-synthesized (dotted line) and photodoped (solid line). The fitting of the experimental data using the multi-layer optical model is depicted by the orange lines. Bottom panel: Numerical simulations of the absorbance of the same ITO-In$_2$O$_3$ NC with $t_s = 4.25$ nm (dotted green lines) and the simulation with extra electrons (*i.e.*, photodoping, solid green line).

By surrounding the core ITO NC with $In_2O_3$, two electronic interfaces are formed: $ITO/In_2O_3$ and $In_2O_3$/surface. In this case, $E_S$ is approximately 0.2 eV below the conduction band minimum of $In_2O_3$, as reported in literature.[2] While in uniform NCs (ITO core only) the band profile is determined by the radial depletion region near the NC surface, the addition of shell layers with thickness $t_s$ strongly affects the band's profile ultimately resulting into a double bending of the conduction band. Hence, shell tuning is an effective tool to control $W$ and the shape of the electronic bands inside the NC volume. This effect can be further extended when combining multiple materials sequentially together in core-multishell NC architectures. **Fig. 1d** reports a heterostructure based on three different materials, introducing three electronic interfaces (other combinations of materials and structures are reported in the Supplementary Information **Fig. S1**). This leads to non-trivial bending and highlights that it is possible to design targeted band structures at the synthesis stage by combining two or more materials in core-shell or core-multishell heterostructures with varying width.

Effective control over NC geometry, size and doping level is crucial to make reliable quantitative assessment of $W$. To experimentally investigate depletion layer engineering predicted by numerical calculations, we synthetized $ITO-In_2O_3$ core-shell NCs with varying shell thickness $t_s$ and induced a dynamic variation of their carrier density via photodoping (see Supplementary Information for further details on synthesis methods). **Fig. 2a** shows the TEM images illustrating the progressive growth of the NC due to the formation of the $In_2O_3$ shell around the ITO core. We collected multiple aliquots during the synthesis at different stages of the growth resulting in a set of samples with the same physical core size ($R_{core}$ = 5.5 nm - C0) and various shell sizes (S1-S5, with $t_s$ = 1.15 nm, 1.9 nm, 2.9 nm, 4.25 nm). The successful achievement of core-shell structures was confirmed by a comparison of the Sn-dopant concentrations obtained by two different techniques: inductively coupled plasma mass spectrometry (ICP-OES) as volume sensitive technique and X-ray Photoelectron Spectroscopy (XPS) as a surface sensitive technique (**Fig. 2b**). These techniques probe the volumetric and surface content of Sn atoms, respectively, and have been shown to be effective methods in elucidating nanocrystal dopant distributions.[25] We observe a higher Sn-concentration from the volume-sensitive measurements (black curve in **Fig. 2b**) as compared to the surface-sensitive measurements (red curve in **Fig. 2b**) in all samples with shell. This indicates that the Sn dopants are localized in the core of the NCs without significant diffusion of Sn atoms into the shell (further analysis on diffusion effects can be found in **Fig. S2**). The absorption spectra of the representative samples, normalized to the maximum are reported in **Fig. 2c** (dotted curves). The spectra are governed by intense resonances in the near-infrared (NIR) that are assigned to localized surface plasmon resonances (LSPRs) as a result of free electrons in the highly doped semiconductor (typically in the range of $10^{27}$ $m^{-3}$ [4,25,26]).[2,27] The LSPR peak position $\omega_{LSPR}$ and its peak profile are correlated to several factors, such as the NC geometrical features (e.g., $R_{core}, t_s$), $n_e$, the depletion layer width ($W$), the dielectric constant of the surrounding medium ($\varepsilon_m$), as well as the structural defects and dopant concentration, providing a unique spectral signature of such parameters.[21,28–30] We will come back to this point later when describing the empirical fit model. From the modulation of the LSPR upon shell growth, we observe an initial blue shift of the LSPR (see Supplementary Information, **Fig. S3**). This is ascribed to the activation of surface dopants with the growth of a thin $In_2O_3$ layer, which results in an increased carrier density.[25] The following continuous red shift of the LSPR is due to the presence of an increasing shell thickness $t_s$ that modifies the dielectric surrounding of the NC.[25] Notably, in particles with a critical thickness $t_s^*$ = 2.7 nm, a second shoulder appears in the spectrum. This indicates a more complex carrier density profile within the core-shell nanocrystals which induces an independent resonating mode, generated by a sufficiently high carrier density in the shell of the nanoparticle.[31]

To further study the electronic structure of core-shell NCs out of the equilibrium conditions, we post-synthetically alter the number of free carriers via photodoping.[4,11,21,32] Photodoping consists of introducing multiple free charge carriers via light absorption and suppressing carrier recombination by the quenching of the holes with hole scavengers.[11,21,33] The photodoping process in colloidal NCs has been recently investigated with optical and electrochemical (e.g., potentiometric titration) measurements.[4,11,21] Here, we induce the photodoping by exposing our colloidal NCs to light beyond the ITO band-gap in the ultraviolet (UV) region (300 nm - 4.1 eV, FWHM = 20 nm) and an intensity of 36.8 mW $cm^{-2}$. **Fig. 2c** shows the normalized absorbance of three representative examples before (dotted curves) and after (continuous curves) the exposure to 20 min of UV light. After the introduction of extra photocarriers into the system the LSPR absorption increases in intensity ($\Delta I = I^{photodope} - I^{as\ synthesize}$) and its energy shifts ($\Delta\omega = \omega_{LSPR}^{photodoped} - \omega_{LSPR}^{as\ synthesized}$). The photo-induced effects progressively appear with the amount of light absorbed (**Fig. S4**) in agreement with previous reports in the literature.[4,16,21,34] The introduced photoelectrons add to the initial free carrier density leading to a stronger interaction with the incoming radiation and hence an increased LSPR absorption. The impact of the photodoping on the LSPR modulation is extremely sensitive to $t_s$, with $\Delta I$ almost doubling in the case of the biggest NCs. In **Fig. 2d**, the normalized absorption spectra for the sample having a $t_s = 4.25$ nm are shown before (black dotted curve) and after (black continue curve) photodoping. In this case it is possible to note a particularly strong splitting of the LSPR into two major contributions. These results display an enhanced sensitivity of the LSPR peak to photodoping by increasing $t_s$ and indirectly hint towards an increased number of stored photoelectrons for higher $t_s$ (since the LSPR absorption is proportional to $n_e^{2/3}$).[19]

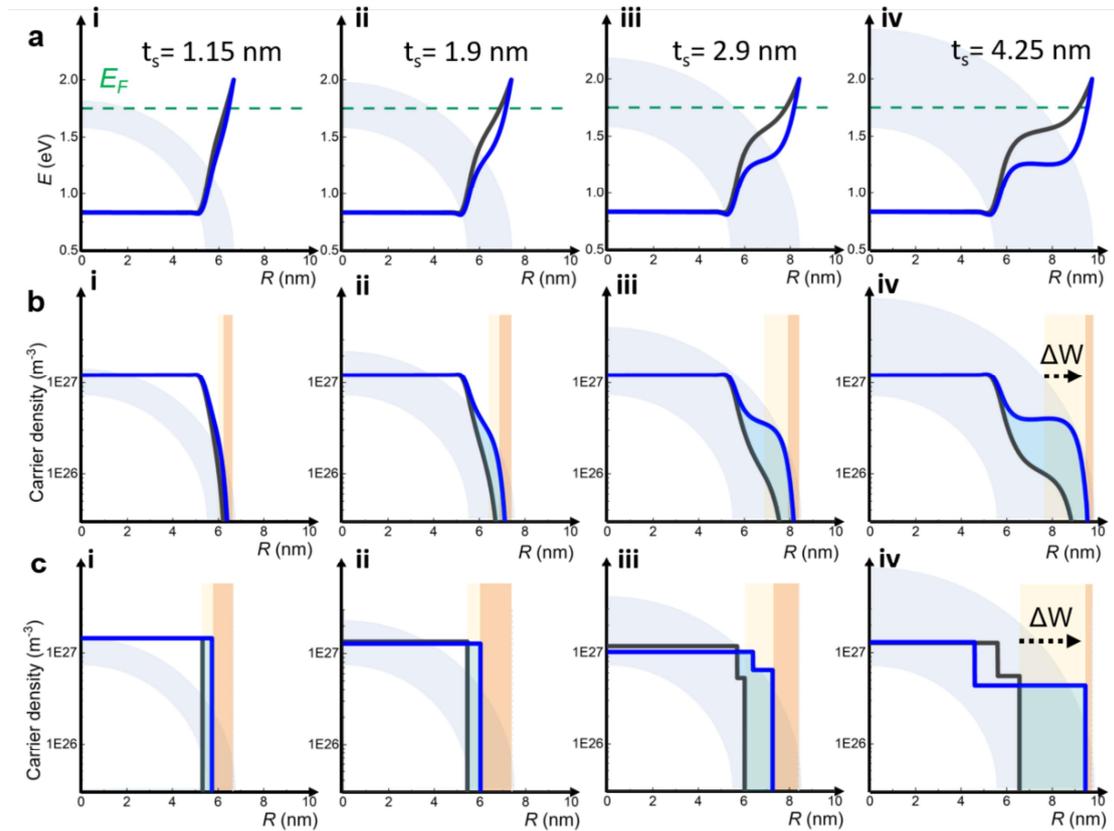

**Fig. 3 | Band profiles and depletion layer modulation via photodoping. a,** Simulated conduction band profiles of core-shell nanocrystals with increasing shell thickness before (black curve) and after (blue curve) the injection of photoelectrons. The light-induced bending of the bands close to the surface of the nanocrystal is the main mechanism responsible for the storage of extra electrons. **b,** Calculated electron density profiles of the same nanocrystals. The depletion region $W$ is shaded in orange. The active region ($R_{active}$) of the NC is observed as the region of the NC volume not affected by $W$. A discrepancy between $R_{active}$ and the structural core ($R_{core}$) and shell ($t_s$) dimensions (given in white and light blue in the background) is observed. The largest variations in $n_e$ after photodoping occur towards the edge region of the nanocrystal with a significant rise of carrier density in the shell region (blue shaded regions). Due to photodoping, the active region expands ($\Delta R_{active} > 0$). Consequently, the depletion layer is progressively suppressed ($\Delta W < 0$) during the storage of electrons with photodoping (from light orange to dark orange). **c,** Carrier density profiles obtained from the fitting of the experimental data using the multi-layer optical model. The expansion of the active region ($R_{active}$) and $W$ modulations of core-shell NCs of increasing $t_s$ upon photodoping are observed similar to from the results from the numerical simulations.

We now investigate the same system of ITO-In$_2$O$_3$ NCs with varying thickness $t_s$ with numerical methods as introduced above. The values for $t_s$ were chosen equivalent to the size of synthesized NCs. To further investigate also the photodoping process, we numerically calculate the effects of additional free electrons in the system as a function of $t_s$. To this aim, we introduced a generation function $G(R) = I_0 \alpha \beta e^{-\alpha R}$, which extends the Poisson's equation by an additional term that represents the spatial distribution of the extra free carriers introduced into the system via photodoping. The intensity of incident photon flux is reported as $I_0$, $\alpha$ denotes the photon absorption coefficient, and $\beta$ denotes the quantum efficiency, respectively.[35] We target to identify how their presence modifies the energy bands and carrier density distribution of the system. In this way, we go beyond the results introduced in **Fig. 1** and we assess the dynamic, post-synthetic variation of the electronic band profiles via light-induced charge injection, *i.e.*, photodoping. A comparative study reporting electronic structures and carrier density profiles both before and after photodoping are shown in **Fig. 3a** and **Fig. 3b**. We first discuss the effects of shell formation on the electronic structure of the NCs (black curves in **Fig. 3a**). The Fermi level pinning anchors the depleted region to the surface of the nanocrystal at the same energy, irrespective of $t_s$. However, with increasing $t_s$ it affects more strongly the In$_2$O$_3$ shell region, which effectively shields the ITO core from depletion. Consequently, even if $W$ increases, the depletion layer progressively shifts towards the outer region of the NC. An intermediate region between core and surface states is, thus, introduced, resulting into the expansion of the active core region ($R_{active}$), *i.e.*, the region of the NC volume not affected by $W$, which is typically larger than $R_{core}$. In fact, the spatial extent of these electronic features does not correspond to the as synthesized structural parameters (*i.e.*, $R_{core}$, $t_s$). This expansion is not due to an introduction of extra donor atoms, nor to diffusion effects (more details in Supplementary Information, **Fig. S2**). With increasing $t_s$, a more pronounced bending of the bands occurs, and it extends for nanometers into the NC. The corresponding carrier density distribution (black curves in **Fig. 3b**) shows a non-trivial profile. The double bending can be explained by a leakage of carriers into the shell. The carrier density in extended regions of the undoped shell reaches values beyond $\sim 1 \cdot 10^{26}$ m$^{-3}$. The presence of carrier density in this range in the undoped In$_2$O$_3$ region indicates that for $t_s > t_s^* = 2.7$ nm it is not appropriate to approximate the ITO-In$_2$O$_3$ system as a doped core-dielectric shell. Instead, it must be considered as a dual-plasmonic material with a specific carrier density in the core ($n_{core} = 1.1 \cdot 10^{27}$ m$^{-3}$) and enhanced carrier density in the shell with $n_{shell} < n_{core}$. This explains the experimentally observed double features in the LSPR (**see Fig. 2c**), which are reproduced by our simulated absorption spectra (**Fig. 2d**) when implementing the carrier density profile extracted from **Fig. 3b**. The observed double bending of the energetic bands becomes more pronounced upon photodoping in all samples (blue curves in **Fig. 3a**) with an immediate impact on the carrier density distribution (**Fig. 3b**). Indeed, after photodoping for samples beyond $t_s^*$ the band profile approaches a step function with two distinct energy levels: the conduction band (CB) level in the core and an energetic level approx. 0.45 eV above the CB in the shell. This effect is observed in the carrier density profile as a (close to) two-step profile. In the samples with $t_s > t_s^* = 2.7$ nm the maximum $n_e$ reached in the NC shell differs from the one in the core, reaching values of around $\sim 4 \cdot 10^{26}$ m$^{-3}$, while the core carrier density remains nearly constant. The light-induced modulation of the depletion layer width ($\Delta W = W^{photodoped} - W^{as\ synthesized}$) increases with a $\Delta W \sim t_s^3$ law (**Fig. S5**). Since the photo-generated extra carriers tend to fill $W$, larger $\Delta W$ values of the as-synthesized NCs justify the possibility to store more electrons in NCs with bigger shells. Hence, from our simulations we conclude that the filling of electronically depleted regions is the main mechanism behind the photodoping process of metal oxide NCs. These findings seem to be in contradiction with the literature reports on the experimentally observed uniform rise of the Fermi energy level as a result of the photo-induced accumulation of multiple photoelectrons, as shown by Schimpf *et al.*.[11,16,21,22] However, the plasmonic effects of double features observed after photodoping of core-shell nanocrystals are not explainable with a simple Fermi level rise. In a flat-band scenario a uniform rise of Fermi level would necessary imply a blue-shift of $\omega_{LSPR}$, while we experimentally observed photodoped NCs with no blue-shift, a red shift (see **Fig. 4b**, below) or even a splitting of the $\omega_{LSPR}$ (**Fig. 2d**) in the core-shell samples. Hence, we hypothesize that the experimentally observed uniform rise of the Fermi level, as observed by Schimpf *et al.*[11,16,21,22] is a result of the averaging over the local carrier densities in different fractions of the NC and in particular in the near surface regions.

To further test our theory, we approach the photodoping process by applying an empirical model to fit the spectra of each sample. Representative fits are shown as orange curves in **Fig. 2d**. In fact, the plasmonic properties of doped MO NCs can be well described within the framework of the Mie scattering theory in the quasi-static approximation (further details in the Supplementary Information). The optical response of metals and heavily-doped semiconductors is characterized by the polarizability of the free electrons, depicted in the Drude-Lorentz model with the complex dielectric permittivity $\varepsilon(\omega) = \varepsilon_\infty - \omega_P^2/(\omega^2 + i\omega\Gamma)$. Here, the bulk plasma frequency $\omega_p = \sqrt{n_e e^2/\varepsilon_0 m^*}$ is a function of the free carrier density ($n_e$) and the effective electron mass ($m^*$), $\Gamma$ is a damping parameter accounting for electron-electron scattering and $\varepsilon_\infty$ is the high frequency dielectric constant. Within this picture, $\omega_{LSPR}$ is directly linked to $\omega_p$ of the material. The tunability of the LSPR is provided by the proportionality to $n_e$, which is related to the number of free charges over the active volume ($n_e \sim N_e/R_{active}^3$). Hence, we can link the absorption, which is our physical observable, to the electronic structure of the system. In previous works, the effect of depletion layers was addressed by Zandi et al., who introduced an effective dielectric function using a Maxwell-Garnett effective medium approximation (EMA).[2] This approach shows that accumulating charges in the NC as a result of electrochemical doping has the effect of increasing the intensity and shifting the position of the LSPR peak as a result of the varying $W$.[2] We adapt this model by implementing a core-shell structure with frequency dependent core dielectric function $\varepsilon_{core}(\omega)$ (and constant carrier density $n_{core}$) surrounded by a dielectric shell with $\varepsilon_{DL} = 4$ in the depletion layer. Outside the sphere a dielectric medium with fixed $\varepsilon_m = 2.09$ is present. Within this picture, we approximate the continuous carrier density profile $n_e(R)$ with discrete regions of uniform density, while we define $n_e = 0$ inside the depletion region (**Fig. 3c**). We found that the two-layer model describes well the optical spectra when $t_s < t_s^* = 2.7\ nm$ (**Fig. 3c**, i and ii). Importantly, for $t_s > t_s^*$ and most photodoping cases, we found that it was necessary to extend this model in order to fit the spectra. To this aim, we developed a three-layer model based on the Maxwell-Garnett EMA with three concentric regions. In the first two regions the inner core and first shell region, which sum up to $R_{active}$, have a frequency dependent dielectric constant of $\varepsilon_{core}(\omega)$ and $\varepsilon_{shell}(\omega)$ with constant carrier densities of $n_{core}(R)$ and $n_{shell}(R)$, respectively. Surrounding the frequency dependent core and shell dielectric functions is an additional layer that accounts for the depletion of carriers in the shell, which was not previously considered in models found in literature.[2,28,31] Hence, these two concentric regions are surrounded by a third depleted layer of thickness $W$ with fixed $\varepsilon_{DL} = 4$ and zero carrier density. The surrounding dielectric medium is $\varepsilon_m = 2.09$. By taking into account the formation of an additional depletion layer due to the electronic interface between shell and surface, our model goes beyond what has been implemented so far to describe capacitive charges in MO NCs.[2,28,31] In our study, for all values of $t_s$, the most notable changes in $n_e$ after photodoping are observed to effectively increase $R_{active}$ and decrease $W$.[28] The core carrier density $n_{core}$ remains nearly constant with variations of less than 14%, while a significant variation occurs in the shell regions, with $n_{shell}$ around $\sim 5.4 \cdot 10^{26}$ m$^{-3}$.

We give a quantitative comparison between the numerical and empirical approach by plotting the amount of stored carriers in the NC ($\Delta N_e$), defined as the difference between the free carriers of the photodoped NC and the as-synthesized NC ($\Delta N_e = N_e^{photodoped} - N_e^{as\ synthesize}$). We observe a good agreement between both approaches, finding that $\Delta N_e$ increases with $t_s$ with a $\Delta N_e \sim t_s^3$ trend, reaching values as high as 600 extra electrons (**Fig. 4a**). We advanced the studies of the NC stored carriers by using titration on photodoped NCs to count the number of stored electrons (further details in the Supplementary Information).[4,11] By using molecular oxidants (F4TCNQ) to titrate the electrons, we directly measure the average number of electrons extracted per NC. F4TCNQ in this study acts as an electron acceptor. The optical features of F4TCNQ hold as a signature to quantify the extracted electrons. We observe an increase of the number of extracted photocarriers with increasing $t_s$, in agreement with the trend reported for numerical simulations and empirical modelling (**Fig. 4a**). The discrepancy, up to a factor 2 in the case of large core-shell NCs, is most probably related to a reduced efficiency in the carrier extraction process. Nevertheless, the $\Delta N_e \sim t_s^3$ trend is reproduced displaying that the electron counting experiments together with the empirical fit model support the band structure calculations well.

Finally, we aim at isolating the impact of depletion layer engineering from the volume dependence of $\Delta N_e$ as shown from Schimpf et al.[16] From numerical simulations, we found that the number of stored carriers in NCs with a core-shell architecture are significantly larger than in the pure ITO case (**Fig. S6a** and **Fig. S6b**). To confirm this result experimentally, we perform a quantitative analysis of one specific $t_s$ and compare it to a similar NC without shell (only core) with all other parameters unchanged (*i.e.,* total NC radius $R$, doping density $N_d$, experimental conditions). The optical absorption spectra before and after photodoping are depicted in **Fig. 4b.** By applying our empirical model to this case, we obtain that core-shell NCs can accumulate ~40% more carriers than uniform ITO NCs of the same size. Our numerical simulations predict that this enhancement increases with increasing shell thickness. Hence, we demonstrate that depletion layer engineering can improve charge storage capacity and more generally that the band bending delivers an additional degree of freedom to artificially engineer the optoelectronic properties of MO NCs, both during synthesis and post-synthetically.

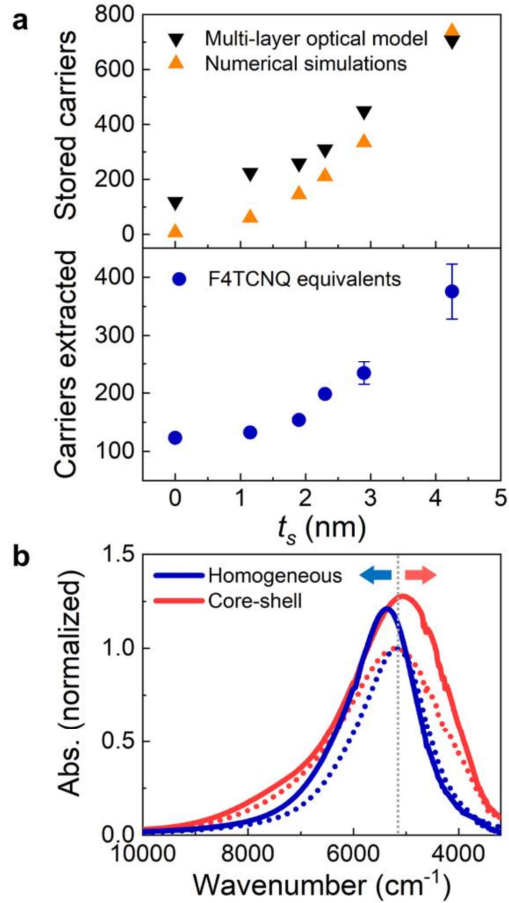

**Fig. 4 | Charge storage enhancement through depletion layer engineering. a,** Upper panel: Photoelectrons stored in the NCs as a function of the shell thickness, showing the $\Delta N_e \sim t_s^3$ trend for values extracted from the empirical multi-layer model (black curve) and the numerical simulations (orange curve). Bottom Panel: Number of electrons extracted via chemical titration using F4TCNQ molecules as a function of $t_s$. **b,** Experimental comparison between the optical response of two samples with same size and doping concentrations but different electronic structure, before (dotted line) and after (continuous line) photodoping (homogeneous ITO in blue, core-shell ITO-$In_2O_3$ in red). The sensitivity of the LSPR modulation via photodoping is enhanced in the core-shell case. We highlight that the peak position of the LSPR after photodoping blueshifts in the homogeneous case, while it redshifts in the core-shell case indicating that depletion layer modulation is the main process of photodoping (see discussion above).

## Conclusions

In this work, we demonstrate that depletion layer engineering is an important tool to design and control energetic band profiles in metal oxide NCs. Our results are based on a combination of theory and experiment: we implement a numerical model that is able to account for additional free carriers in the MO NCs, we developed an empirical three-layer model that describes the optical response of the (photodoped) core-shell ITO-$In_2O_3$ NCs, and confirmed our results with experimental electron counting experiments through reaction with F4TCNQ. From this combined theoretical and experimental approach, we found that, first, double bending of the bands dominates the electronic structure of (photodoped) core-shell ITO-$In_2O_3$ NCs and that the depletion layer predominantly affects the $In_2O_3$ shell. Second, the electronic rearrangement of energy bands and the filling of electronically depleted regions resulting in the evolution of different levels of carrier density in core and shell are the main mechanism behind the photodoping process of metal oxides NCs. Third, depletion layer engineering allows enhancing the charge storage capability of ITO NCs of the same size. We can extend this model to other systems as well demonstrating the validity of our approach. Our results show that the modulation of the depletion layer represents an interesting avenue to design and improve the properties of MO NCs and their core-shell or core multishell structures. We foresee multiple practical applications ranging from energy storage to sensing for ITO- and similar metal oxide nanocrystals-based devices that will benefit from the control of electronic band profiles through depletion layer engineering.

## Acknowledgements

For this work, the authors acknowledge the support of both European Union's Horizon 2020 European Research Council, under grant agreement no. 850875 (Light-DYNAMO), and European Union's Horizon 2020 Research and Innovation program under grant agreement no. 101017821 (LIGHT-CAP).

## Author contribution

M.G. and N.C. contributed equally to this work. M.G. designed the experiment, synthesized and characterized the nanomaterial, performed photodoping and titration experiments, developed the multi-layer optical model, and wrote the manuscript. N.C. coordinated the research work, performed numerical simulations, analyzed the experimental data and wrote the manuscript. M.B.L. performed numerical simulations, N.P. developed the multi-layer optical model. M.W. performed titration experiments, M.P. performed XPS measurements and analyzed the data. A.F. and L.M. supervised the research. I.K. designed the experiment, supervised and coordinated all research work and wrote the manuscript.

## Competing financial interests

The authors declare no competing financial interest.

## Methods

**Core-Shell Nanocrystal synthesis**
ITO/In$_2$O$_3$ core/shell nanocrystals were synthesized with the following procedure, adapted from Ref. [19,25,36]: a precursor solution was prepared mixing in a flask tin(IV) acetate and indium(III) acetate in a 1:9 Sn:In ratio. Subsequently, oleic acid was added in a 1:6 metal to acid ratio to yield a 10% Sn doped ITO precursor solution. The flask was left at 150 $^0$C under N$_2$ for 3 hours. The ITO nanocrystals (core) were first prepared by adding the ITO precursor solution via a syringe pump (drop-by-drop at a rate of 0.35 mL/min) to 13.0 mL of oleyl alcohol at 290 $^0$C. During the slow-injection procedure a flow of 130 mL/min of N$_2$ gas was kept in the reaction flask. The ITO cores were grown to a size of 5.5 nm (radius) and isolated by precipitating with ~12 mL ethanol. The solid part was collected by centrifugation at 7300 rpm for 10 min, washed twice more with ethanol and dispersed in hexane.

Then, part of the cores was kept for analysis and the rest of the solution reintroduced in fresh oleyl alcohol. For shelling, a second precursor solution was prepared in an analogue manner. In order to yield an undoped indium oleate precursor solution, indium(III) acetate was mixed with oleic acid in a 1:6 molar ratio. Undoped indium oleate was added with the same slow-injection procedure described above. Core-shell samples were washed with ethanol and the process repeated several times until a final size of ~10 nm was reached. All experiments were performed on samples collected at different stages of the shell growth, and hence sharing the very same ITO core.

**Structural characterization of core-shell NCs**
The structural characterization of the samples with different shell thickness were analyzed by transmission electron microscopy (**TEM**) to determine the size and confirm the successful formation of nanocrystals. TEM measurements were performed with a JEOL JEM-1400Plus operating at 120 kV and using lacey carbon grids supported by a copper mesh. The size distribution of the NCs was extracted using ImageJ tools on the images collected.[37]

X-ray Diffraction (**XRD**) analyses were carried out on a PANanalytical Empyrean X-ray diffractometer equipped with a 1.8 kW Cu Kα ceramic X-ray tube and a PIXcel3D 2x2 area detector, operating at 45 kV and 40 mA. Specimens for the XRD measurements were prepared by dropping a concentrated NCs solution onto a zero-diffraction silicon substrate. The diffraction patterns were collected under ambient conditions using a parallel beam geometry and the symmetric reflection mode. XRD data analysis was carried out using the HighScore 4.1 software from PANalytical.

X-ray Photoemission Spectroscopy (**XPS**) measurements were performed on a Kratos Axis Ultra$^{DLD}$ spectrometer, using a monochromatic Al Kα source (15 kV, 20 mA). Specimens were prepared by dropping a concentrated NCs solution onto a highly ordered pyrolytic graphite (HOPG, ZYA grade) substrate. High resolution spectra of the Sn 3d and In 3d regions were acquired at pass energy of 10 eV, and energy step of 0.1 eV, over a 300 x 700 microns area. The photoelectrons were detected at a take-off angle of $\phi = 0°$ with respect to the surface normal. The pressure in the analysis chamber was maintained below $7 \times 10^{-9}$ Torr for data acquisition. The data was converted to the VAMAS format and processed using the CasaXPS software, version 2.3.24.[38] The binding energy (BE) scale was internally referenced to C 1s peak (BE for C–C = 284.8 eV). For the quantitative analysis, the areas of In 3d and Sn 3d peaks were calculated after applying the appropriate background correction across the binding energy range of the peaks of interest. The relative atomic concentrations were then estimated, using the so-called relative sensitivity factors (RSF) provided by Kratos ($RSF_{In\ 3d} = 7.265$, $RSF_{Sn\ 3d} = 7.875$).

Inductively coupled plasma mass spectrometry (**ICP-OES**) was performed on all samples to estimate the doping levels and concentrations of the ITO NCs. The elemental analysis was carried out via inductively coupled plasma optical emission spectroscopy (ICP–OES) on an iCAP 6000 Series ICP–OES spectrometer (Thermo Scientific). In a volumetric flask, each sample was dissolved in aqua regia [HCl/HNO$_3$ 3:1 (v/v)] and left overnight at RT to completely digest the NCs. Afterward, Milli-Q grade water (18.3 MΩ cm) was added to the sample. The solution was then filtered using a polytetrafluorethylene membrane filter with 0.45 μm pore size. All chemical analyses performed by ICP–OES were affected by a systematic error of about 5%.

**Optical measurements**
Optical measurements were carried out on a Cary5000 UV−vis−NIR Spectrometer. Spectra were collected in anhydrous toluene in the range 280-3200 nm with a scan resolution of 1 nm. After drying the solvent, ITO NCs were transferred in anhydrous toluene (Sigma-Aldrich) in a nitrogen filled glove box. Rectangular anaerobic cuvettes with a sealed screw cap (Starna Scientific) were used for photodoping and titration experiments.

**Photodoping process**
Before photodoping, the ITO-In$_2$O$_3$ NCs were dissolved in anhydrous toluene, as described above. Subsequently, the photodoping process on the NCs is achieved by illuminating the quartz cuvette containing the solution with the NCs with a UV LED (central wavelength: 300 nm, bandwidth: 20 nm). The cuvette was placed at a distance of 12 mm from the cuvette window (Thorlabs M300L4). UV power density at the front window of the cuvette was 36.8 mW cm$^{-2}$.

**Redox titration**
The titrant was prepared by dissolving 0.34 mg of F4TCNQ (2,3,5,6-tetrafluoro-7,7,8,8-tetracyanoquinodimethane) in 30 mL of anhydrous toluene. The titrant addition steps were carried out in the inert environment of a nitrogen-filled glove box to avoid any contact with ambient oxygen. Electron counting was performed after photodoping by spectroscopic analysis of the neutral, anion and dianion forms of the F4TCNQ molecules. The effects of titrants on as-synthesized ITO-In$_2$O$_3$ samples were tested, showing no sign of interaction in the spectrum.

**Multi-layers fitting model for LSPRs**

The distinct dielectric response of the core-shell NCs is implemented as an effective dielectric function $\varepsilon_{eff}(\omega)$ based on the Maxwell-Garnett effective medium approximation (EMA). This model is further extended to consider multiple shell regions and corresponding dielectric environments. We fit the experimental data with a particles warm optimization algorithm in MATLAB (R2020a. Natick, Massachusetts: The MathWorks Inc.) and we extract the carrier densities $n_{e,core}$ and $n_{e,shell}$ and spatial extensions ($R_{core}$, $R_{shell}$) of the core and shell regions, respectively, for each NC of increasing shell thickness before and after photodoping.

**COMSOL Simulations**

Simulations for the energy band and carrier density profiles were solved numerically for spherical NCs using a finite-element method. Poisson's equation was solved with the software COMSOL Multiphysics v5.0 (Comsol Inc., Burlinghton MA USA) using a finite-element scheme (see Supplementary Information for details).

# Supplementary Information

# Control of electronic band profiles through depletion layer engineering in core-shell nanocrystals

Michele Ghini[†,1,2] Nicola Curreli[†,3*] Matteo B. Lodi,[4] Nicolò Petrini,[3] Mengjiao Wang,[3,5] Mirko Prato,[5] Alessandro Fanti,[4] Liberato Manna,[1] Ilka Kriegel[3*]

1 - Department of Nanochemistry, Istituto Italiano di Tecnologia, via Morego 30, 16163 Genova, Italy
2 - Dipartimento di Chimica e Chimica Industriale, Università degli Studi di Genova, Via Dodecaneso 31, 16146 Genova, Italy
3 - Functional Nanosystems, Istituto Italiano di Tecnologia (IIT), via Morego 30, 16163 Genova, Italy
4 - Department of Electrical and Electronic Engineering, University of Cagliari, via Marengo 2, 09123, Cagliari, Italy
5 - Materials Characterization Facility, Istituto Italiano di Tecnologia, Via Morego 30, 16163, Genova, Italy
† - These authors contributed equally to this work.

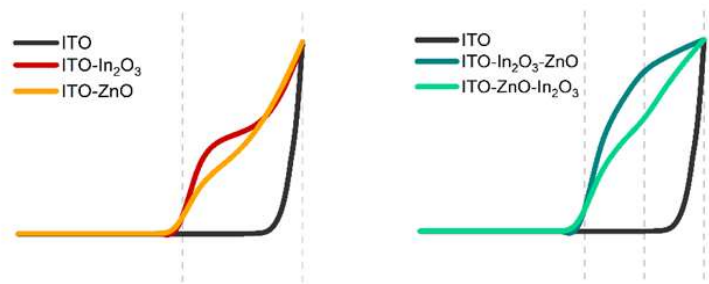

**Fig. S1 | Depletion layer engineering of metal oxide NCs via tuning of structural and electrical properties.** Effects of multiple shells of different materials surrounding an ITO core radius of 5.5 nm. A homogeneous ITO NCs of the same total radius (total radius = 9.5 nm) is shown in black, as comparison. **a,** ITO-$In_2O_3$ core-shell and ITO-ZnO core-shell NCs. **b,** ITO-$In_2O_3$-ZnO core-multishell and ITO-ZnO-$In_2O_3$ core-multishell NCs. Each shell has a thickness of 2 nm, for a total NC radius of 9.5 nm.

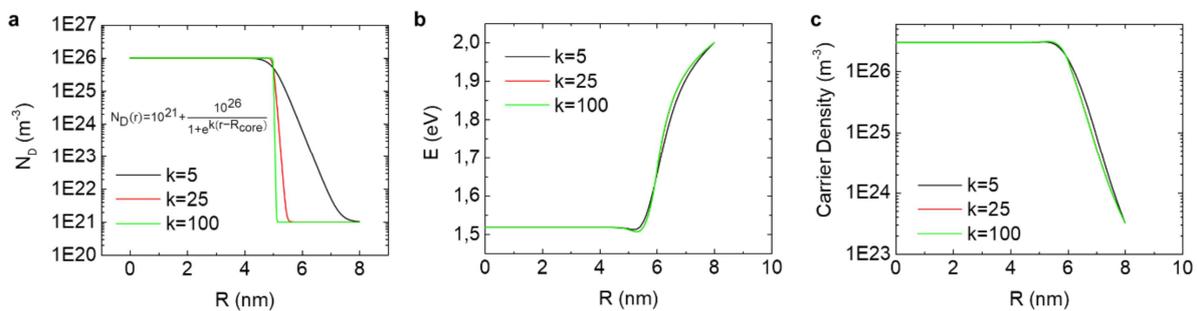

**Fig. S2 | Simulated effects of dopants diffusion.** By varying the shape factor ($k = 5, 25, 100$), we induce a variation of the dopant distribution within the core-shell nanocrystal. **a,** The Poisson's equation was solved, and the band diagram and electron density were analysed finding that differences in the conduction band and electron density profiles are negligible. **b,** The numerically calculated energy levels and **c,** the numerically calculated carrier density profiles of an ITO-$In_2O_3$ core-shell NC with the donor profiles input as shown in panel **a**.

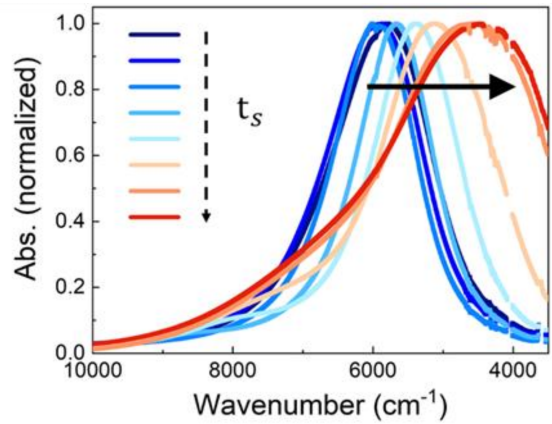

**Fig. S3 | Normalized absorption spectra of ITO-In$_2$O$_3$ core-shell samples.** By growing a subnanometric shell the peak position of the LPSR gets blueshifted, due to the activation of dopants at the surface of the ITO core. Further growth of undoped shell has the effect of red-shifting the main peak of the LSPR in the core-shell system.

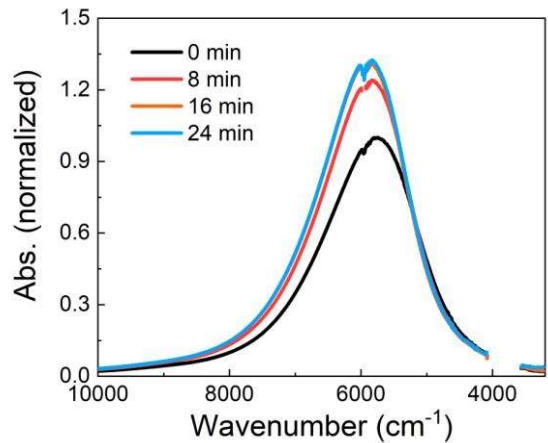

**Fig. S4 | Photodoping ITO-In$_2$O$_3$ core-shell NCs.** Temporal evolution ($t = 0, 8, 16, 24$ min of UV exposure) of the LSPR absorption lineshape during the photodoping process. Most of the variations occurs in the first minutes, as expected for the charging of a capacitor-like system. After 16 minutes of UV exposure photodoping saturates, with little modification of the LSPR lineshape.

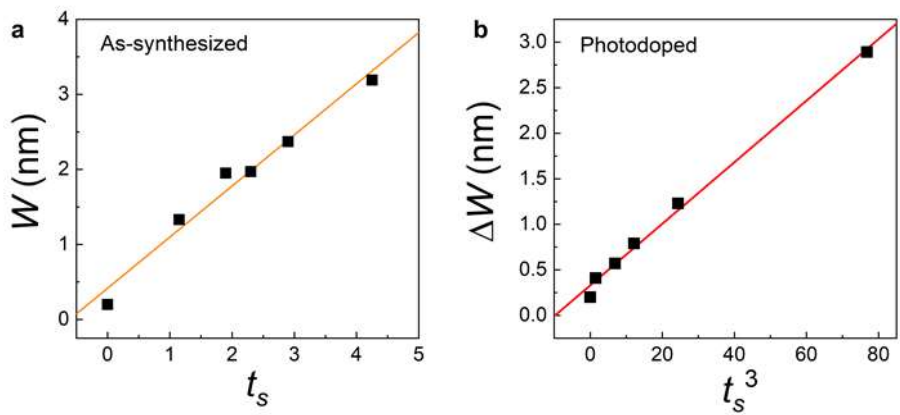

**Fig. S5 | Dependence of main parameters on the shell thickness. a,** Depletion layer width as a function of the shell thickness $t_s$. A linear fit is reported in orange. **b,** The contraction of the depletion layer width ($\Delta W$) in core-shell ITO-In$_2$O$_3$ NCs follows a $t_s^3$ law (linear fit showed in red), with $t_s$ being the shell thickness.

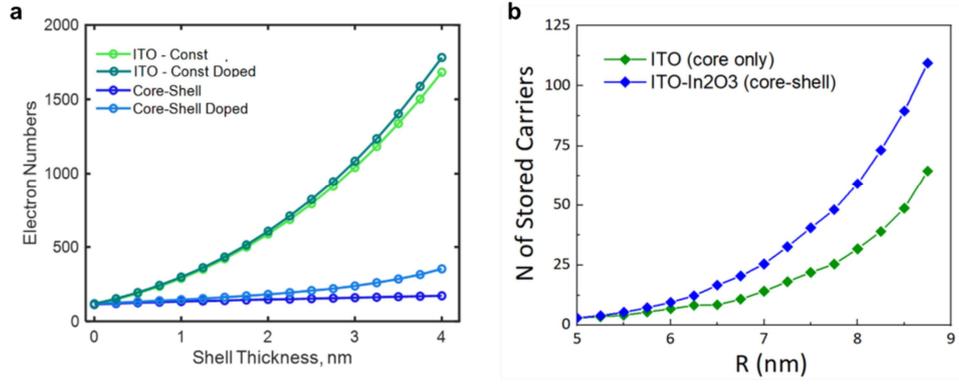

**Fig. S6 | Comparison between the core-shell architecture and homogeneous particles. a,** Numerical calculations of the total number of free carriers in the two cases of a homogeneous NC (ITO – core only) and core-shell NC (ITO-In$_2$O$_3$), as a function of the shell thickness before and after photodoping. **b,** Number of photocarriers stored in the two cases as a function of the radius.

## Numerical calculations details

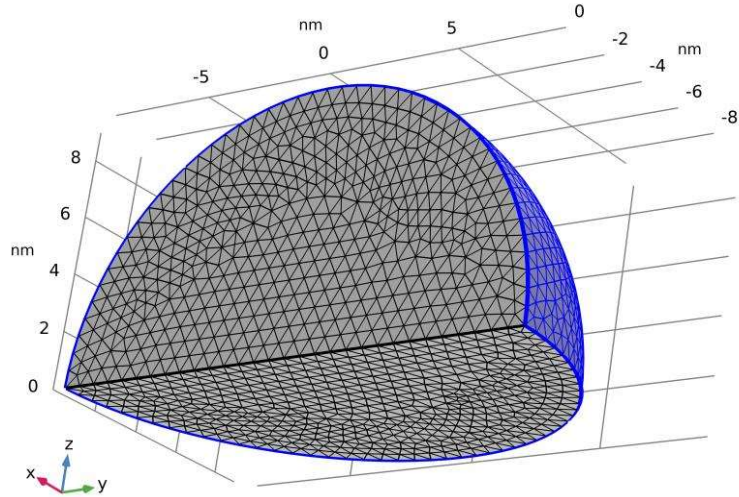

**Fig. S7 | Meshed geometry of a spherical nanocrystal.** Poisson's equation was solved over this control volume using a finite element method in COMSOL.

For a spherical nanocrystal (Fig. S7) with given radial dopant profile, $N_D$ and surface potential, $E_S$, we solved numerically the Poisson's equation using a finite element method (FEM) software Comsol Multiphysics v5.0 (Comsol Inc., Burlinghton MA USA) by using the built-in *Mathematics module* exploiting the central symmetry of the sphere as shown in Fig. S7.

We adapted the dimensionless form of Poisson's equation in Cartesian coordinates derived by Seiwatz and Green:[1–3]

$$\nabla^2 u = -\frac{\rho \cdot e^2}{\varepsilon \varepsilon_0 k_B T} \tag{1}$$

where $e$ is the electron charge, $k_B$ is the Boltzmann constant, $T$ is temperature, $\varepsilon_0$ is the vacuum permittivity, $\varepsilon$ is the static dielectric constant, and $\rho$ is the charge density. The non-dimensional potential, $u$ refers to the difference between the neutral bulk and the surface potentials. It is defined as:[1]

$$u = u_{bul} - u_{surface} = \frac{E_F - E_I}{k_B T} \tag{2}$$

where $E_F$ is the Fermi energy level, and $E_I = \frac{E_{CB} + E_{VB}}{2}$ is the reference potential and center of the band gap in which $E_{CB}$ and $E_{VB}$ are the conduction band and valence band profile respectively.

It is possible to expand the Poisson's equation by defining the charge density, $\rho = \rho(r)$ given by the following relation:

$$\rho = \rho_D(r) + p - n \tag{3}$$

where $\rho_D(r)$ is the radially changing donor dopant density, while $p$ and $n$ are the contribution of the hole and electron density, respectively.

Using an auxiliary function $w_{x,y} = \frac{E_x - E_y}{k_B T}$ it is possible to further define the terms of the charge density. Given the donor energy level, $E_D$, the activated dopant concentration can be expressed as:

$$\rho_D(r) = \frac{N_D(r)}{1 + 2e^{(u - w_{D,I}(r))}} \tag{4}$$

where $w_{D,I} = \frac{E_D - E_I}{k_B T}$.

The free hole ($p$) and electron ($n$) concentration in the parabolic conduction band are equal to:

$$p = 4\pi \left[\frac{2m_h k_B T}{h^2}\right]^{3/2} F_{1/2}(w_{V,I}(r) - u);$$

$$n = 4\pi \left[\frac{2m_e k_B T}{h^2}\right]^{3/2} F_{1/2}(u - w_{C,I}(r)) \tag{5}$$

where $w_{V,I} = \frac{E_{VB} - E_I}{k_B T}$, $w_{C,I} = \frac{E_{CB} - E_I}{k_B T}$ and $h$ is the Planck's constant.

Here, the Fermi-Dirac distribution is assumed for the carriers:

$$F_{\frac{1}{2}}(\eta) = \int_0^\infty \frac{x^{1/2} dx}{1 + e^{x - \eta}} \tag{6}$$

being $x$ the dummy variable, and the upper integration limit is set to ∞ on account of the fact that the integrand vanishes exponentially at high energies. From a numerical point of view, the upper bound of the integral was selected to be a finite number high enough to ensure the convergence to a solution, and to retain an accuracy within the 1%.

The Poisson's equation become:

$$\nabla^2 u = -\frac{e^2}{\varepsilon\varepsilon_0 k_B T}\left\{\frac{N_D(r)}{1+2e^{(u-w_{D,I}(r))}} + 4\pi\left[\frac{2m_h k_B T}{h^2}\right]^{3/2} F_{1/2}(w_{V,I}(r) - u) - 4\pi\left[\frac{2m_e k_B T}{h^2}\right]^{3/2} F_{1/2}(u - w_{C,I}(r))\right\} \quad (7)$$

where $m_h$ is the effective hole mass equal to 0.6 times the electron mass and $m_e$ is the effective electron mass equal to 0.4 times the electron mass.

**Parameters used in the simulations**

**Tab. 1 | Parameters used in the simulations.**

| $e$ | $\varepsilon$ | $k_B$ | $T$ | $m_h$ | $m_e$ | $h$ |
|---|---|---|---|---|---|---|
| 1.602 $10^{-19}$ C | 8.854 $10^{-12}$ Fm$^{-1}$ | 1.381 $10^{-23}$ JK$^{-1}$ | 293.15 K | 5.466 $10^{-31}$ kg | 3.644 $10^{-31}$ kg | 6.626 $10^{-34}$ Js |

For the simulations we used the parameters introduced in the Tab. 1. In the case of nanocrystals made of only one material, energy levels, number of donors and relative dielectric permittivity are constant.

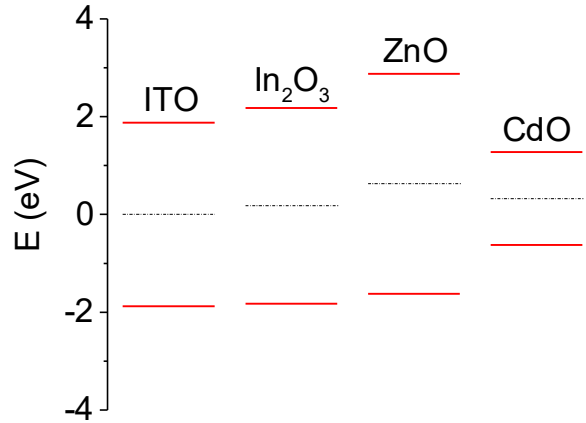

**Fig. S8 | Relative conduction and valence band levels for ITO, In$_2$O$_3$, ZnO and CdO nanocrystals.** The energy levels are referred as non-equilibrium states. Dashed lines represent the intrinsic level.

For systems made of combination of different materials, since the NCs are material-dependent, we must define the radius-dependent non-equilibrium potentials profiles, donor distribution profile ($N_D = N_D(r)$) and relative dielectric profile ($\varepsilon(r)$).

For example, in the case of ITO-In$_2$O$_3$ nanocrystals, the radius-dependent non-equilibrium potentials profiles and donor distribution profile used to solve the Poisson's equation are shown in Fig. S9.[4] Different distributions can be found for ZnO and CdO (Fig. S8).[5–7]

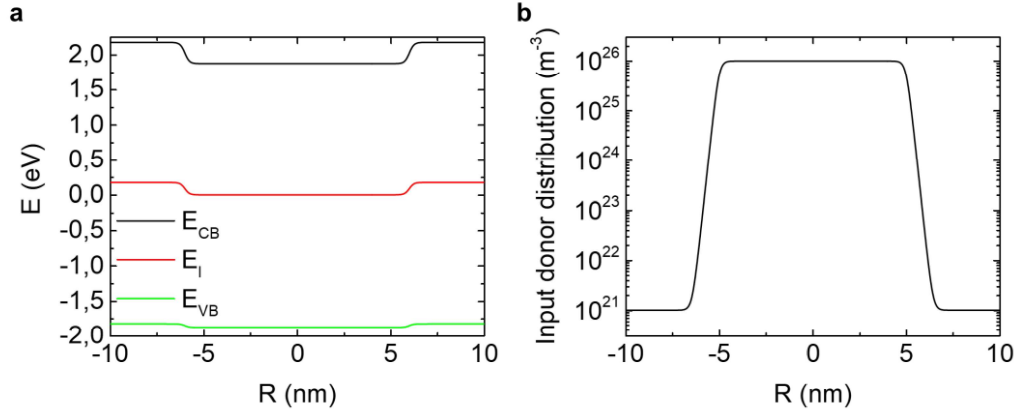

**Fig. S9 | Potential and donor distribution in ITO-In$_2$O$_3$ nanocrystal with 6 nm core and a 5 nm shell. a,** Input band diagram for a core-shell ITO- In$_2$O$_3$ structure. The system reference is the intrinsic level of the core material. **b,** Input donor distribution as a function of the nanocrystal radius.

The diffusion effects on the NCs are investigated by modifying the donor distribution profile $N_D(r)$ (Fig. S2). In detail, we used a sigmoid-like function to model the diffusion effects:

$$N_D(r) = 10^{21} + \frac{10^{26}}{1+e^{k(r-r_{core})}} \tag{8}$$

By varying the shape factor ($k = 5, 25, 100$), the Poisson's equation was solved, and the band diagram and electron density were analysed again finding that differences in the conduction band and electron density profiles are negligible.

For each system the energy levels and carrier density profiles after equilibrium were obtained according to the specific material taken into account and their combinations. The results can be seen in Fig. 1 of the main text and in Fig. S1.

**Photodoping**

To study the core-shell system after photodoping we introduced additional generation and recombination terms into the Poisson's equation:[8]

$$\nabla^2 u = -\frac{\rho \cdot e^2}{\varepsilon \varepsilon_0 k_B T} + G(r) - R(r) \tag{9}$$

The recombination effects are neglected, hence $R(r) = 0$. In particular, the generation term $G(r)$ is spatially dependent and is modelled by the following Gaussian distribution:

$$G(r) = \frac{N_{max}}{V} \frac{1}{\sigma\sqrt{2\pi}} e^{\frac{r-R_{NC}}{\sigma}} \tag{10}$$

where $N_{max}$ is the maximum and peak number of photoelectrons and $\sigma$ is the shape factor which determine the spatial distribution of the photoelectrons into the nanocrystal. By varying $\sigma$ we are able to model the effects on the energy levels after the photodoping and its effect on the band profiles. The results are shown in Fig. 2 of the main text.

**Electrons stored**

The total number of electrons in the NC is calculated by integrating the electron density over the particle volume ($V$):

$$N_E = \int_V n(r)dV \tag{11}$$

The trend found by numerical simulations is in accordance with the experimental findings (Fig. 4 main text). The simulation tends to underestimate the number of stored electrons for small $t_s$, however the error is within the margin of uncertainty of this method. We computed the number of stored electrons for all cases, and the results are shown in Fig. 4 of the main text.

**Optical simulations**

In Fig. S10 and Fig. 2d of the main text, we show the numerical simulations of absorption spectrum of ITO-In$_2$O$_3$ NCs. We couple the solution based on the radial carrier distribution obtained from the Poisson's equation, with the high-frequency formulation of the Maxwell's equation relating the electronic density of the NCs to the plasmonic resonance and simulating the absorption spectra. The simulations are carried out in Comsol by using the *Electromagnetic Waves, Frequency Domain module* adding a background medium box 10 times bigger than the nanocrystal radius to compute the far-field contribution.

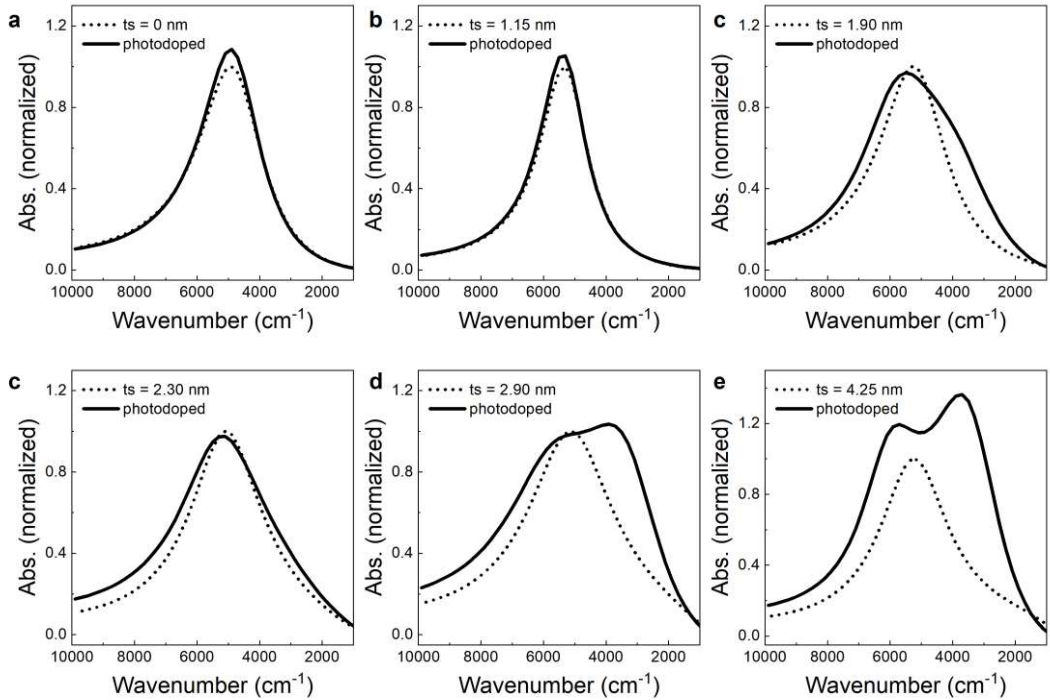

**Fig. S10 | Numerical simulations of the absorbance of the ITO-In$_2$O$_3$ core-shell NC** of different shell thickness ($t_s$) before (dotted line) and the after the addition of extra electrons (*i.e.*, photodoping, solid line). Values of $t_s$ were selected in order to match the samples C0-S5 we extensively analyzed.

**Optical model**

We estimate the number of electrons accumulated via photodoping from the experimental absorption spectra of ITO-In$_2$O$_3$ NCs by implementing an optical model based on Mie solution and effective medium approximation. Briefly, we considered Mie solution under the hypothesis of spherical particles in the quasi-static limit, and assuming a negligible retardation effect. Moreover, the effect of scattering is also negligible and can be disregarded. Under those conditions, the absorption relation can be written as:[9]

$$ABS = \frac{\sigma_{abs} N L}{\ln(10)} \tag{12}$$

where $\sigma_{abs}$ is the far-field absorption cross-section, $N$ is the volume density of particles, $L$ is the light pathlength through the cuvette. Considering the hypothesis of dominant dipolar mode,[10] absorption cross-section can be calculated as:

$$\sigma_{abs} = 4\pi k_m R^3 Im\left\{\frac{\varepsilon - \varepsilon_m}{\varepsilon + 2\varepsilon_m}\right\} \tag{13}$$

where $k_m = \frac{2\pi}{\lambda}\sqrt{\varepsilon_m}$ is the wavevector in the medium $m$, and $R$ is the particle radius, $\varepsilon$ is the dielectric permittivity of the nanoparticle and $\varepsilon_m$ is the permittivity of the medium. For a homogeneous particle, the dielectric permittivity $\varepsilon$ can be defined in the framework of the free-electron Drude-Lorentz model for bulk metals and doped semiconductor:

$$\varepsilon(\omega) = \varepsilon_\infty - \omega_P^2/(\omega^2 + i\omega\Gamma) \tag{14}$$

where $\varepsilon_\infty$ is the high frequency dielectric permittivity, $\omega_p$ is the bulk plasma frequency and $\Gamma$ is a damping parameter accounting for electron-electron scattering. In the case of ITO NCs, the damping parameter is described by an empirical frequency-dependent damping function:

$$\Gamma(\omega) = \Gamma_L - \frac{\Gamma_L - \Gamma_H}{\pi}\left[\tan^{-1}\left(\frac{\Gamma_X}{\Gamma_W}\right) + \frac{\pi}{2}\right] \tag{15}$$

where $\Gamma_L$ and $\Gamma_H$ are the low and high frequency dumping constants, $\Gamma_X$ is the crossover frequency in the mixed regime and $\Gamma_W$ is the width of the crossover region. Frequency-dependent damping function accounts the electrons scattering of ionized impurities and allows accurately reproducing the asymmetry of the plasmonic resonances of ITO NCs.[11,12]

In the Equation (14), the bulk plasma frequency, $\omega_p$ is a function of the free carrier density ($n_e$) and the effective electron mass ($m^*$):

$$\omega_p = \sqrt{\frac{n_e e^2}{\varepsilon_0 m^*}} \tag{16}$$

where $e$ is the electron charge and $\varepsilon_0$ is the vacuum permittivity. Therein, from calculating the plasma frequency it is possible to obtain the carrier density of a homogeneous nanoparticle. In the case of the heterogeneous nanoparticle, the expression above must take into account the different materials and their geometrical arrangement inside the spherical particle. The Maxwell-Garnett effective medium approximation can successfully be applied in order to obtain an effective dielectric constant $\varepsilon_{eff}(\omega)$ which is used in the studied case of core-shell structures.[13] The mixing formula yields:

$$\varepsilon_{eff}(\omega) = \varepsilon_{shell}\left(\frac{(\varepsilon_{core}+2\varepsilon_{shell})+2F(\varepsilon_{core}-\varepsilon_{shell})}{(\varepsilon_{core}+2\varepsilon_{shell})-F(\varepsilon_{core}-\varepsilon_{shell})}\right) \qquad (17)$$

where $\varepsilon_{eff}(\omega)$ is the effective dielectric constant, $\varepsilon_{shell}$ is the dielectric permittivity of the shell material, $\varepsilon_{core}$ is the dielectric permittivity of the core, $F = (R_{core}/R)^3$ is the volume ratio between the core volume and the total nanoparticle volume. Both $\varepsilon_{shell}$ and $\varepsilon_{core}$ are calculated with Equation (14). Equation (17) can be applied recursively in order to calculate multiple layered core-shell structures. In particular, we exploited a three-layer structure for modelling the depletion layer-shell-core arrangement inside the studied NCs (Fig. S11). The effective dielectric constant is obtained as:

$$\varepsilon_{eff1}(\omega) = \varepsilon_{shell}\left(\frac{(\varepsilon_{core}+2\varepsilon_{shell})+2F_1(\varepsilon_{core}-\varepsilon_{shell})}{(\varepsilon_{core}+2\varepsilon_{shell})-F_1(\varepsilon_{core}-\varepsilon_{shell})}\right);$$

$$\varepsilon_{eff}(\omega) = \varepsilon_{DL}\left(\frac{(\varepsilon_{eff1}+2\varepsilon_{DL})+2F_2(\varepsilon_{eff1}-\varepsilon_{DL})}{(\varepsilon_{eff1}+2\varepsilon_{DL})-F_2(\varepsilon_{eff1}-\varepsilon_{DL})}\right) \qquad (18)$$

where $\varepsilon_{eff1}(\omega)$ is the effective dielectric constant considering only the core-shell structure without the depletion layer, $F_1 = (R_{core}/(R_{core}+t_s))^3$ is the volume ratio between the core volume and the core + shell volume, $F_2 = ((R_{core}+t_s)/R)^3$ is volume ratio between the core + shell volume and the total nanoparticle volume, $\varepsilon_{DL}$ is the dielectric permittivity of the depletion layer. In the case of the depletion layer permittivity $\varepsilon_{DL}$, Equation (14) simplifies to $\varepsilon_{DL}(\omega) = \varepsilon_\infty$ since the plasma frequency of the depletion layer $\omega_{p,DL} = 0$ (since $n_e = 0$).

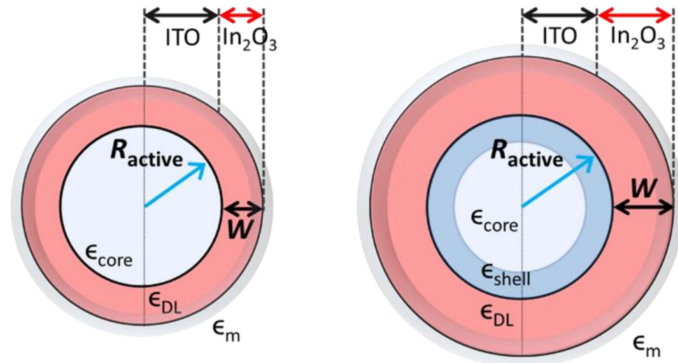

**Fig. S11 | Multi-layer model for core-shell plasmonic NCs.** Schematic illustration of the multi-layer model implemented to analyse the optical response of ITO-In$_2$O$_3$ NCs. The inner part of the volume delimited with $R_{active}$, is assumed to be the only region in which the electrons are free to oscillate and contribute to the plasmon resonance. This active region consists of an inner uniformly doped core with dielectric constant $\varepsilon_{core}$ and a shell with dielectric constant $\varepsilon_{shell}$, initially completely depleted of free electrons, responsible for a second mode of the LSPR. The two concentric spheres are surrounded by a depleted layer (with fixed $\varepsilon_{DL}$) of thickness $W$ and they are immersed in a dielectric medium (fixed $\varepsilon_m$).

**Fitting model**

In order to fit the spectra, a three-layer model was implemented. While a two-layer model (structured with one active region and one superficially depleted region) is sufficient to successfully fit as-synthesised and photodoped spectra of small particles, it fails for particles with a shell thickness ($t_s$) larger than the critical thickness of $t_s^* = 2.7$ nm. This choice relies on the fact that in the case of one peak spectrum only one electronic core oscillator is needed, while when in core-shell NCs a double-peak absorption appears two oscillators are needed. The two-layer model can be seen as a particular case of the three-layer model. As a matter of fact, when the contribution from the shell is negligible a single electronic population can be considered responsible for the plasmonic response. The remaining part of the nanoparticle is just occupied by the depletion layer, which is plasmonically inactive but alters the dielectric environment that the core experiences. In contrast, the three-layer model was needed to correctly fit all samples. Some NCs developed a double peak response only after photodoping (due to the addition of extra charges in the shell), while even bigger NCs presented the double feature even in the as-synthesised case. In this latter case, two of the three layers are plasmonically active (core and shell), while the third layer is held by the depletion region. Moreover, the observed redshifts of the main peak in the absorption spectra after photodoping require the presence of the depletion layer to be explained for both the aforementioned cases.

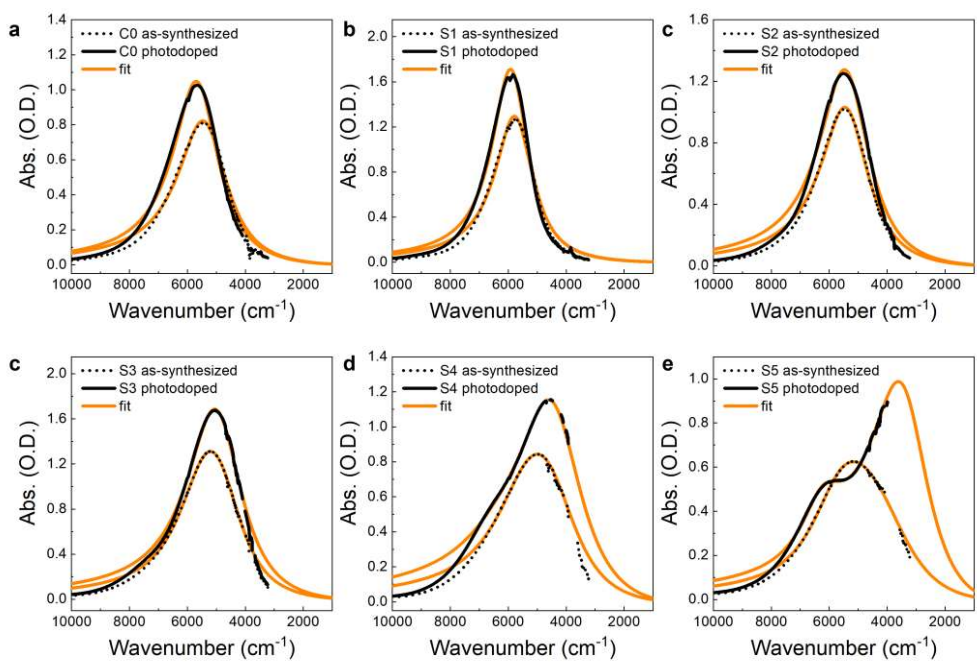

**Fig. S12 | Absorbance of sample C0-S5** (with values of shell thickness from zero to 4.25 nm) in the as-synthesized (dotted line) and photodoped (continue line) cases. The fitting of the experimental data using the multi-layer optical model is reported with the orange lines.

The two-layer model relies on the fit of six parameters, which are the number of electrons in the core, $n_{e,core}$, the number of electrons in the shell, $n_{e,shell}$, the core radius, $R_{core}$, the core damping parameter, $\gamma_c$, the shell damping parameter, $\gamma_s$, and the solution volumetric concentration, $p = N_p \cdot V_p$, where $N_p$, is the density of nanoparticle in the solution and $V_p$, is the average nanoparticle's volume. The three-layer model adds one more parameter to the previous ones: $S$, which is the radius of the active shell. The algorithm used for finding the best fit with both the models is the particle swarm optimization algorithm, implemented with Matlab software. This is preferable to least squares fitting function because the solution does not depend on the initial guess of the parameters but the global best fit (if exists) is found exploring all the possible combination of parameters between the lower and higher bounds. It must be noticed that the bounds for the photodoped case depends on the solution of the same as-synthesised particle. In order to have consistent solutions without overfitting, a final comparison between the solutions found for the as-synthetized case with the photodoped ones is needed. The best–fit found for the as-synthesized and photodoped samples are showed in Fig. S.12, and the extracted parameters are reported in Tab. 2 and Tab. 3 respectively.

**Tab. 2| Extracted parameters for three-layer model fit for the as-synthetized samples.**

| As-synthetized | C0 | S1 | S2 | S3 | S4 | S5 |
|---|---|---|---|---|---|---|
| $N_{e,core}$ | 703 | 911 | 912 | 916 | 914 | 937 |
| $N_{e,shell}$ | 0 | 0 | 0 | 49 | 75 | 250 |
| $R_{core}$ (nm) | 5.3 | 5.32 | 5.45 | 5.62 | 5.7 | 5.6 |
| $R_{active}$ (nm) | 5.3 | 5.32 | 5.45 | 5.83 | 6.03 | 6.56 |
| $\gamma_c$ (cm$^{-1}$) | 2.08 | 1.62 | 2.04 | 1.88 | 2.35 | 1.66 |
| $\gamma_s$ (cm$^{-1}$) | 10 | 10 | 10 | 3.19 | 3.62 | 4.60 |
| $p$ | 3.40e-5 | 5.68e-5 | 7.87e-5 | 11.98e-5 | 11.73e-5 | 11.71e-5 |
| $W$ (nm) | 0.2 | 1.33 | 1.95 | 1.97 | 2.37 | 3.19 |

**Tab. 3| Extracted parameters for three-layer model fit for the photodoped samples.**

| Photodoped | C0 | S1 | S2 | S3 | S4 | S5 |
|---|---|---|---|---|---|---|
| $N_{e,core}$ | 823 | 1136 | 1171 | 1177 | 1111 | 528 |
| $N_{e,shell}$ | 0 | 0 | 0 | 99 | 327 | 1365 |
| $R_{core}$ (nm) | 5.5 | 5.73 | 6.02 | 6.38 | 6.39 | 4.59 |
| $R_{active}$ (nm) | 5.5 | 5.73 | 6.02 | 6.62 | 7.26 | 9.45 |
| $\gamma_c$ (cm$^{-1}$) | 1.91 | 1.53 | 2.12 | 2.01 | 2.87 | 1.36 |
| $\gamma_s$ (cm$^{-1}$) | 10 | 10 | 1.97 | 5.67 | 2.60 | 2.58 |
| $p$ | 3.4024e-5 | 5.68e-5 | 7.87e-5 | 11.98e-5 | 11.73e-5 | 11.71e-5 |
| $W$ (nm) | 0 | 0.92 | 1.38 | 1.18 | 1.14 | 0.3 |